\documentclass[aps,prl,amsmath,amssymb,twocolumn,floatfix,10pt,showpacs]{revtex4-1}
\usepackage{graphicx}
\usepackage{hyperref}
\usepackage{color}
\usepackage{SIunits}
\usepackage[normalem]{ulem}

\newcommand{\ve}{\mathbf}

\begin{document}

\title{Quantum Nature of Edge Magnetism in Graphene}
\author{Michael Golor, Stefan Wessel, and Manuel J. Schmidt}
\affiliation{Institut f\"ur Theoretische Festk\"orperphysik, JARA-FIT and JARA-HPC, RWTH Aachen University, 52056 Aachen, Germany}
\date{\today}


\begin{abstract}
It is argued that the subtle crossover from decoherence-dominated classical magnetism to fluctuation-dominated quantum magnetism is experimentally accessible in graphene nanoribbons. We show that the width of a nanoribbon determines whether the edge magnetism is on the classical side, on the quantum side, or in between. In the classical regime, decoherence is dominant and leads to static spin polarizations at the ribbon edges, which are well described by mean-field theories. The quantum Zeno effect is identified as the basic mechanism which is responsible for the spin polarization and thereby enables the application of graphene in spintronics. On the quantum side, however, the spin polarization is destroyed by dynamical processes. The great tunability of graphene magnetism thus offers a viable route for the study of the quantum-classical crossover.

\end{abstract}

\maketitle

Electron-electron interactions in solid-state systems, containing large numbers of electrons, are exceedingly difficult to treat and thus pose one of the greatest challenges for theoretical physics. This is especially true for interactions that are too strong to be accounted for perturbatively, i.e., the regime of {\it strong correlations} with its huge variety of exotic phases. Graphene is usually not considered to be a strongly correlated material \cite{kotov_interactions_review_2012}, because the vanishing density of states (DOS) at the charge-neutrality point suppresses magnetic correlation effects very efficiently. However, the DOS only vanishes in a perfect bulk crystal. Imperfections, such as edges or vacancies give rise to additional electronic states at the Fermi level \cite{fujita_1996}. They result in a peak in the local DOS, with the striking consequence that these imperfections enter the regime of strong correlations. 

The central phenomenon in this context is the so-called edge magnetism (EM) \cite{golor_armchair_2013,lee_edge_magnetism_2005,yazyev_chiral_edge_magnetism_2011,luitz_ed_2011,karimi_2012,son_abinitio_prl_2006,jung_edge_magnetism_2009,wakabayashi_edge_magnetism_1998,golor_chiral_qmc_2013,li_prl_2013}, which is discussed as having possible applications in spintronics \cite{yazyev_mag_corr_2008,son_half_metallic_gnrs_2007}. The simplest geometry for EM is a nanoribbon with perfect zigzag edges. In this case, all theories that we are aware of, predict (or are usually interpreted to the effect of) an extended spin polarization along the edges in the ground state, with opposite spin directions at opposite edges. One might call this a non-local N\'eel state in the sense that the opposite spins are not neighbors on atomic distances, but are spatially separated. Moreover, this N\'eel state is implicitly assumed to be classical, i.e., non-fluctuating, just as the well-known Heisenberg antiferromagnet. This picture is rooted in the often-used mean-field approaches to EM, such as Hartree-Fock or {\it ab-initio} methods, which are the simplest methods for the treatment of strong electronic interactions. They approximate a problem of interacting fermions by a problem of non-interacting fermions, complemented by a self-consistency condition. But they disregard quantum fluctuations and, in the present context, break the $SU(2)$ symmetry of the initial problem in an uncontrolled way. More elaborate approaches, such as quantum Monte-Carlo (QMC) \cite{feldner_prl_2011,feldner_prb_2010,golor_chiral_qmc_2013} and density matrix renormalization group \cite{hikihara_2003}, have been applied to EM, but these methods are restricted to rather small systems in which at least the static spin correlations agree well with mean-field results \cite{feldner_prb_2010,feldner_prl_2011}. Thus, the mean-field picture of EM with static spin polarizations prevailed in the community.

\begin{figure}
\centering
\includegraphics[width=\columnwidth]{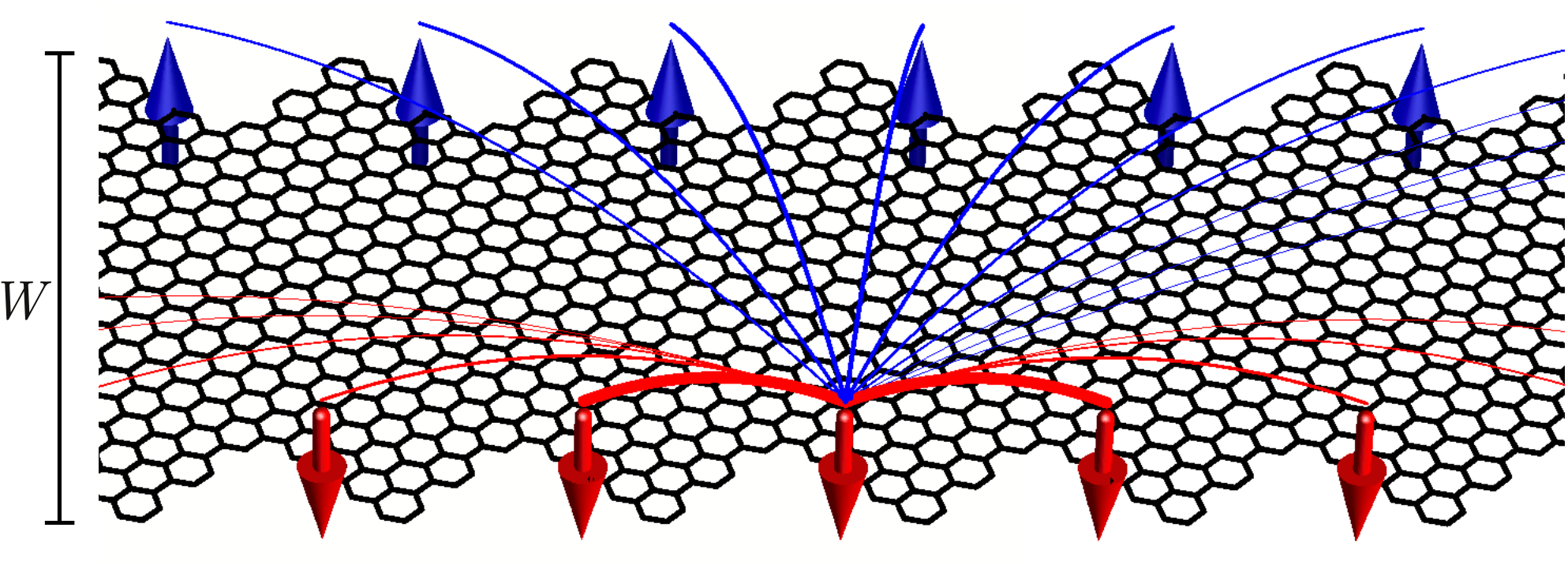}
\caption{Special ribbon geometry which allows for a controlled mapping to a spin-$\frac12$ quantum Heisenberg model with a single spin located on each zigzag segment. The effective spin-spin interactions are ferromagnetic (antiferromagnetic) along (across) the edges and sketched here for one reference spin.}
\label{fig_geometry}
\end{figure}

In this Letter, we use a recently developed method \cite{schmidt_eff_vs_qmc_2013} allowing us to study EM in realistically large systems without the above-mentioned mean-field artefacts. The central idea of this method is the derivation of an effective quantum Heisenberg theory for the edge states. For special edge geometries in which the edge states are well localized and separated from each other (see Fig.~\ref{fig_geometry}), this effective theory has been shown to be in quantitative agreement with numerically exact QMC methods \cite{schmidt_eff_vs_qmc_2013}. We restrict our quantitative calculations to this geometry, but our qualitative arguments are expected to extend to more general geometries, including the pure zigzag edge and those chiral edges for which magnetic features are expected \cite{yazyev_chiral_edge_magnetism_2011}.

We will discuss the different classical and quantum aspects of EM and their interrelation -- the quantum-classical crossover -- via the quantum Zeno effect \cite{misra_quantum_zeno_1977}. Our arguments and calculations are based on extreme geometries and limits in which exact calculations become feasible. We thus aim at establishing a way of thinking about edge magnetism, which is not based on mean-field theory and which is consistent with rigorous theorems \cite{lieb_theorem} and results from exact methods \cite{feldner_prl_2011,feldner_prb_2010,golor_chiral_qmc_2013,hikihara_2003}.

{\it Models and methods.} Our work is based on the lattice Hubbard model for graphene $H_{\rm l} = -t \sum_{\langle i,j\rangle,\tau} c^\dagger_{i\tau} c^{}_{j\tau} + U \sum_i c^\dagger_{i\uparrow} c^{}_{i\uparrow} c^\dagger_{i\downarrow}c^{}_{i\downarrow}$, where $c_{i\tau}$ annihilates an electron with spin $\tau$ at site $i$. $\langle i,j\rangle$ runs over nearest neighbors. In the spectrum of $H_0$ there are two eigenstates $\phi_{k\pm}$ with energies $\pm \epsilon_k$ smaller than those of all other eigenstates for a given momentum $k$ along the edge. These states are the (anti-)symmetric combinations of the $A$ and $B$ edge states $\phi_{k\pm}(i) = \phi_{kA}(i)\pm \phi_{kB}(i)$, from which we can reconstruct the actual edge states $\phi_{ks}(i)$, where $s=A,B$ labels the edge.

Following Ref.~\onlinecite{schmidt_eff_vs_qmc_2013}, we construct a Wannier basis $\phi_{xs}(i) =L^{-1/2} \sum_{k}e^{-i k x + i\varphi_k} \phi_{ks}(i)$, where $x$ enumerates the zigzag segment at each edge, $L$ is the number of zigzag segments along the edge, and $\varphi_k$ are phases, optimized numerically such that $\sum_i|\phi_{xs}(i)|^4$ is maximal. It is convenient to characterize the ribbons in terms of zigzag segments $L$ and the width $W$ in nm. $L$ is also the number of effective spins at each edge. The essential approximation in the derivation of the Heisenberg theory is the assumption that each Wannier state is occupied by exactly one electron. This is well justified in our geometry \cite{schmidt_eff_vs_qmc_2013}. The electron spins in the Wannier states are described by the Heisenberg model
\begin{equation}
H_{\rm H} = \sum_{x,x'} J^{\rm AF}_{xx'} \ve s_{xA}\cdot \ve s_{x'B} - \sum_{s,x<x'} J^{\rm FM}_{xx'} \ve s_{xs}\cdot\ve s_{x's},\label{ham_heis}
\end{equation}
where $\ve s_{x s}$ are vectors of spin-$\frac12$ operators. The coupling constants are given by $J^{\rm FM}_{xx'}= U \sum_{i} |\phi_{xs}(i)|^2|\phi_{x's}(i)|^2$ and $J^{\rm AF}_{xx'} = 4|t_{xx'}^*|^2/( U \sum_{i}|\phi_{xs}(i)|^4)$, with the effective inter-edge hopping $t^*_{xx'} = -t \sum_{\langle i,j\rangle} \phi^*_{xA}(i)\phi_{x'B}(j)$. The antiferromagnetic terms $J^{\rm AF}_{xx'}$ result from the usual combination of hopping between electronic states $t_{xx'}$ and an on-site repulsion $\propto U$. The ferromagnetic $J^{\rm FM}_{xx'}$, however, is mediated by a direct wave function overlap of Wannier states via the Hubbard Hamiltonian, which prefers parallel spin alignments. For more details, see Ref. \onlinecite{schmidt_eff_vs_qmc_2013}. The couplings in $H_{\rm H}$ are not restricted to nearest neighbors, but spread over distances that depend on the ribbon width $W$. The typical decay length (in unit cells) of $J^{\rm FM}_{xx'}$ is $\xi_{\rm FM} \approx 0.203 \frac{W}{\text{nm}} + 0.078 \frac{W^2}{\text{nm}^2}$ and that of $J_{xx'}^{\rm AF}$ is $\xi_{\rm AF} \approx  0.57\frac{W}{\text{nm}} -0.14$.
$H_{\rm H}$ is thus reminiscent of a Heisenberg ladder with ferromagnetic (antiferromagnetic) leg (rung) couplings, which are, unlike in conventional Heisenberg ladders, smeared out over many neighboring spins. The notion of Heisenberg ladders in EM is not new (cf. Refs.~\onlinecite{wakabayashi_edge_magnetism_1998,yazyev_mag_corr_2008,yoshioka_ribbonladder_2003}), but we present here a controlled microscopic derivation and analyze its full quantum nature.

Note that $H_{\rm H}$ is not frustrated and is thus accessible by QMC methods. Here, we employed the stochastic series expansion (SSE) method~\cite{sandvik_longrange_2003,fukui_walkeralias_2009}, which is based on a high-temperature series expansion of the partition function, and enables us to calculate spin-spin correlations, gaps and susceptibilities of spin systems over a large temperature range.

\begin{figure}
\centering
\includegraphics[width=\columnwidth]{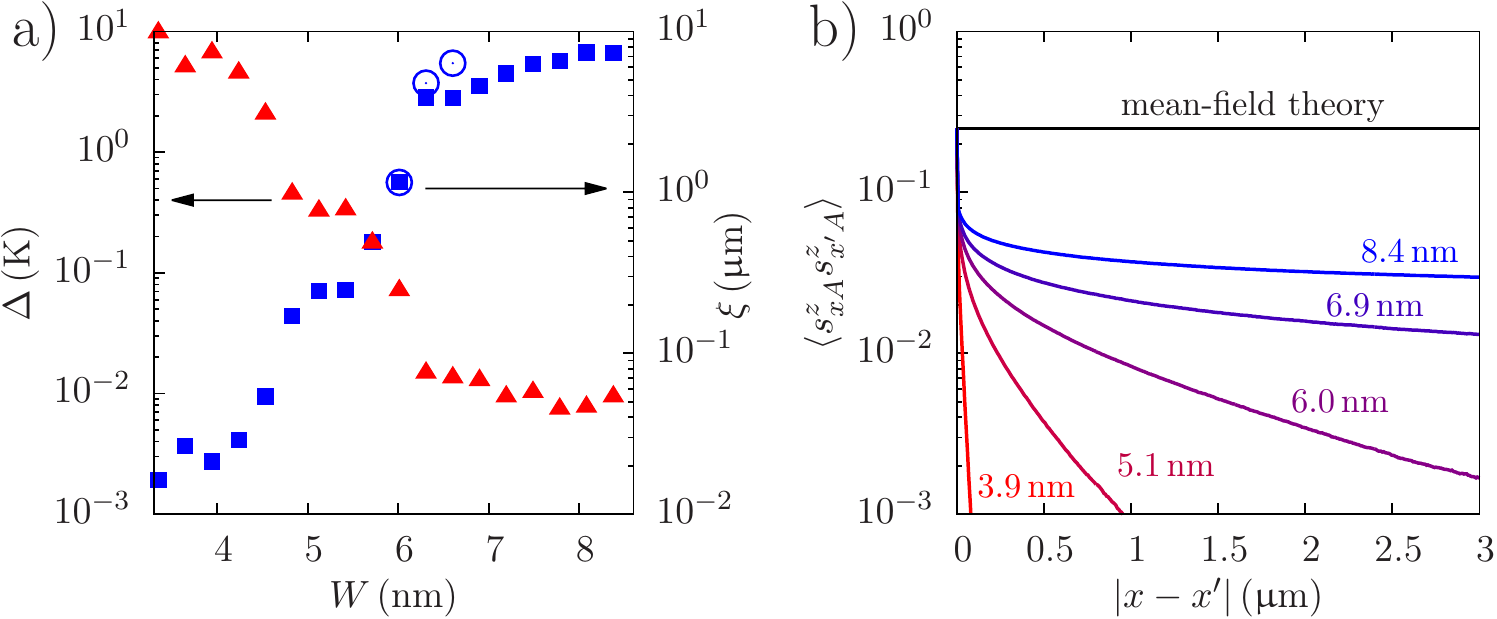}
\caption{(Color online) (a) Spin-spin correlation length $\xi$ and spin gap $\Delta$ for different widths $W$, calculated for $12\,\micro\meter$ long ribbons at temperature $T=10\,$mK (errors are smaller than symbol size). Open circles show results for twice as long ribbons (i.e., $24\,\micro\meter$). (b) Spin-spin correlations $\langle s^z_{xA} s^z_{x'A}\rangle$ along the edge for different widths $W$.  For comparison, the black line shows the constant polarization as expected from mean-field theory.}
\label{fig_correlations}
\end{figure}

{\it Magnetic edge correlations and spin gap.} Figure \ref{fig_correlations} shows the intra-edge spin correlations as well as the spin gap for ribbons of different widths $W$. In our QMC simulations we mostly used ribbons with size $L=8000$, which corresponds to $16000$ edge spins and a physical ribbon length of about $12\,\micro\meter$. 
We perform all simulations at a temperature $T=10\,$mK, which we verified to be sufficiently low to obtain ground state correlations for ribbons with width at least up to $W\sim 6.0\,$nm.
The distance over which the spins are correlated along the same edge grows rapidly with $W$. We extract the correlation length $\xi$ from the spin correlations by fitting an exponential function to the asymptotic tails, 
 $\langle s^z_{xA}s^z_{x'A}\rangle\propto\exp(-|x-x'|/\xi)$. From the analogy with standard Heisenberg ladders with ferromagnetic leg couplings it is expected that $\xi$ grows exponentially with $W$ \cite{Minoru_class_2d_fm_87,Mikeska_ferro_ladder_96}, which is in qualitative agreement with our results. The deviations from the exponential behavior of $\xi$ for $W\gtrsim 6\,$nm is a finite size effect, as the total length of the ribbon we simulate is on the order of $\xi$. Note, however, that we actually consider realistic ribbon lengths, so that this finite size effect is by no means an artefact but an experimentally relevant regime. Results for even longer ribbons ($24\,\micro\meter$) support further exponential growth of the correlation length. It should also be noted that mean-field theory predicts infinite $\xi$ at zero temperature in all ribbons considered here.

The spin gap $\Delta$ is estimated by performing simulations at different temperatures (always remaining close to the ground state) and then fitting the expected low-temperature behavior $\chi(T\rightarrow 0)\propto\exp(-\Delta/T)/\sqrt{T}$ to the obtained uniform susceptibilities~\cite{troyer_spingap_1994}. The spin gap behaves, as expected, inversely to $\xi$, i.e. it decreases with $W$, including the finite size effect for $W\gtrsim6$ nm. It is remarkable that the spin gap is tunable over more than two orders of magnitude via a moderate change in $W$ roughly by a factor of three. Such small spin gaps are below the resolution of conventional fermionic QMC techniques.

Furthermore, it should be noted that the rugged behavior of $\Delta$ and $\xi$ as functions of $W$ is not due to statistical/numerical errors (the error bars are smaller than the symbols used in Fig. \ref{fig_correlations}), but has its origin in the special geometry we use here (see Fig. \ref{fig_geometry}). Due to the shifting of the relative spatial spin positions at different edges \footnote{The ribbons we use are actually composed of unit cells that are straight narrow armchair ribbons with two zigzag ends (each hosting a spin). These are connected to each other along the armchair edges, but tilted relative to each other. From Fig. 1 one may convince oneself that increasing $W$ shifts the blue spins to the right of the red spins.}, the inter-edge coupling is not smooth in $W$.

{\it Quantum dynamics.} For the discussion of the dynamical aspects of EM we define the total edge spins $\ve S_s = \sum_{x} \ve s_{xs}$, the total spin $\ve S = \ve S_A+\ve S_B$, the operator for the staggered magnetic moment $\ve S_{\rm st} = \ve S_A-\ve S_B$, and the Hamiltonian for an artificial symmetry breaking field $H_{\rm st} = - h_{\rm st} S_{\rm st}^z$. Due to the invariance of $H_{\rm H}$ under spin rotations, $\ve S^2=S(S+1)$ is a good quantum number.

It is known that the ground state $|\Psi_0\rangle$ of a finite-size quantum Heisenberg antiferromagnet, such as $H_{\rm H}$, is a spin singlet ($S=0$). This is in perfect consistence with Lieb's theorem \cite{lieb_theorem}, which asserts that the ground state of $H_{\rm l}$ is a singlet as well. However, the thermodynamic limit (TL) of $|\Psi_0\rangle$ is problematic. The staggered magnetization is zero $\langle\Psi_0| \ve S_{\rm st}|\Psi_0\rangle=0$, but its fluctuations $\langle \Psi_0|\ve S_{\rm st}^2|\Psi_0\rangle\sim L^2$ are extensive. In the context of antiferromagnetism, such a state is called non-ergodic, and is not considered as a valid ground state in the TL \cite{koma_afm_finite_size_1994}.

In contrast, a classical N\'eel-like state $|\Psi_{\rm N}\rangle$ with opposite spin polarizations at opposite edges is a more reasonable candidate for a ground state of conventional antiferromagnetic systems. It has an extensive staggered magnetization with non-extensive fluctuations. In other words, $|\Psi_{\rm N}\rangle$ is ergodic (i.e., behaves classically) and admissible as a ground state for antiferromagnets in the TL. But $|\Psi_{\rm N}\rangle$ is not an eigenstate for finite-sized systems. These two opposing viewpoints have been discussed extensively for ordinary antiferromagnets and their unification is now well understood (see, e.g., Ref.~\onlinecite{koma_afm_finite_size_1994}). Essentially, a {\it tower of excited states} with $S=0,1,\dots$, from which $|\Psi_{\rm N}\rangle$ is formed, collapses in the TL and forms the macroscopic ground-state degeneracy, which is needed for the spontaneous time-reversal and $SU(2)$ symmetry breaking \cite{Anderson_52,Neuberger_89}. The remainder of this Letter is concerned with the implementation  of this principle in EM.

In particular, we discuss how the two opposing viewpoints, (1) $|\Psi_0\rangle$ is a ground state but non-ergodic and (2) $|\Psi_{\rm N}\rangle$ is ergodic but not an eigenstate, can be reconciled from a quantum-dynamical perspective in the context of EM. We define the N\'eel state $|\Psi_{\rm N}\rangle$ for our Heisenberg theory as the ground state of $H_{\rm H} + H_{\rm st}$ for $h_{\rm st}$ chosen such that $\langle \Psi_{\rm N} | S^z_{\rm st} |\Psi_{\rm N} \rangle/L = 0.95$ \footnote{Note that the exact choice of this initial polarization is not important for our conclusions.}. For conventional antiferromagnets in the TL, the corresponding $h_{\rm st}$ approaches zero and the usual definition $\lim_{h_{\rm st}\rightarrow0}\lim_{L\rightarrow\infty} |\Psi(h_{\rm st})\rangle$ of the symmetry-broken ground state emerges.

As discussed above, $|\Psi_{\rm N}\rangle$ is not an eigenstate of $H_{\rm H}$ and will therefore decay on a certain time scale $\tau_{\rm qd}$ (for reasons, which will become clear below, 'qd' stands for quantum decay). For ordinary antiferromagnets, $\tau_{\rm qd}$ diverges in the TL. For EM, we extract $\tau_{\rm qd}$ by analyzing the quench dynamics of the staggered magnetic moment
\begin{equation}
D_{\rm st} (t) = L^{-1} \langle \Psi(t) | S^z_{\rm st} |\Psi(t)\rangle,
\end{equation}
where $|\Psi(t)\rangle = \exp(-i t H_{\rm H}) |\Psi_{\rm N}\rangle$. For general ribbon lengths $L$ and widths $W$ it is very difficult to calculate $D_{\rm st}$. For ribbons with a high aspect ratio ($L \ll W$), however, the inter-edge antiferromagnetic couplings are essentially independent of $x-x'$. If the ribbon width $W$ is twice its length $L$ we have $\xi_{\rm AF} \approx 1.7 L$, so that the antiferromagnetic couplings are approximately constant $J^{\rm AF}_{xx'} \approx J^{\rm AF} = (5/L)^3$~K. Thus, as far as the antiferromagnetic part of $H_{\rm H}$ is concerned, $H_{\rm H}$ is equal to the exactly solvable Lieb-Mattis (LM) model of antiferromagnetism \cite{lieb_mattis_afm_1962}. We need to perform a further approximation in order to be able to avail ourselves of the exact solution of the Lieb-Mattis model, namely the assumption of constant ferromagnetic intra-edge coupling. From the size scaling of $J^{\rm FM}_{xx'}$, this assumption is not justifiable. However, we will see that, making this assumption $J^{\rm FM}_{xx'} = J^{\rm FM}$, the results are completely independent of $J^{\rm FM}$, since all spins at the same edge behave as one large superspin. Thus, we do not expect this approximation to affect our results in a qualitative way.

The spectrum of the LM-approximated $H_{\rm H}$ is $E = J^{\rm AF} S(S+1)/2- (J^{\rm AF}+J^{\rm FM}) \sum_s S_s(S_s+1)/2$ \cite{lieb_mattis_afm_1962}. Thus, in the ground state we have $S=0$ and $S_s=S_{\rm max}=L/2$. The crucial simplification due to the LM approximation lies in the fact that $S_s$ is invariant under $H_{\rm st}$. Thus, it is sufficient to expand $H_{\rm st}$ in the basis $|S\rangle = |S,S_A,S_B\rangle$ with $S_s=S_{\rm max}$. One finds $H_{\rm st} |S\rangle = 2 h_{\rm st} L (a_S |S-1\rangle + a_{S+1}|S+1\rangle)$, where the coefficients $a_S$ are defined recursively in Ref.~\onlinecite{kaplan_sym_break_afm_1990}, and $S = 0,1,\dots, L$. The full Hamiltonian $H_{\rm H}+ H_{\rm st}$, projected to the relevant subspace, is thus an $(L+1)$-dimensional matrix, which can easily be diagonalized numerically for very large systems up to $L\sim 10^4$. The spectrum of the LM-approximated $H_{\rm H}$ gives rise to a further simplification: In the relevant sector $S_A=S_B=S_{\rm max}$ all excitation energies are multiples of the first excitation energy $J^{\rm AF}$. Thus, the quench dynamics show a full revival after $t_r \sim (J^{\rm AF})^{-1}$, which provides us with an upper bound for $\tau_{\rm qd}$.

\begin{figure}
\centering
\includegraphics[width=\linewidth]{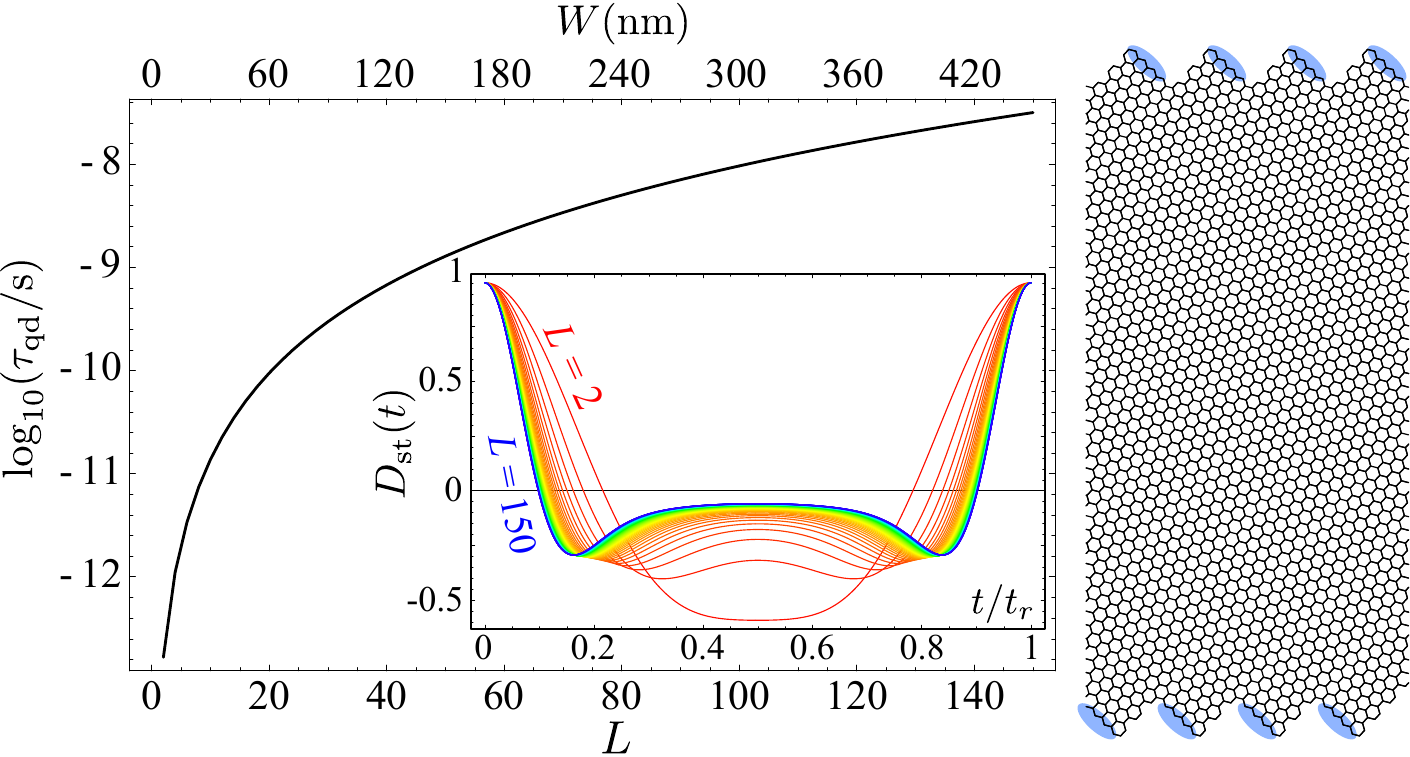}
\caption{(Color online) Life time $\tau_{\rm qd}$ of the N\'eel state in the Lieb-Mattis approximation for different ribbon dimensions. $L$ is the number of effective edge spins along the edge. For convenience, the width $W$ of the ribbon is given in nm in the upper abscissa scale. On the right side an example ribbon with aspect ratio 2 ($L=4$ unit cells; $W=12$ nm) is shown. For different $L$ the ribbon's aspect ratio is invariant. Periodic boundary conditions in $x$-direction are assumed. The edge spins are indicated as blue ellipses. The inset shows the exact time evolution $D_{\rm st}(t)$ for $L=2,4,6,\dots,150$. The time axis is rescaled by the revival time $t_{\rm r} = 2 \pi / J^{\rm AF}$.}
\label{fig_lm_lifetime}
\end{figure}

Figure \ref{fig_lm_lifetime} shows $D_{\rm st}(t)$ for different ribbon sizes $L$ with a fixed aspect ratio of 2. We define the decay time as the first zero of $D_{\rm st}(t)$. Apparently, $\tau_{\rm qd} \sim 0.1 t_{\rm r}$ over a wide range of system sizes $L$. Note that this simple relation between decay time and revival time is special to the LM approximation. In general, the excitation spectrum is incommensurate and thus the revival time is, as usual, exponentially large in the system size. Thus, we have obtained a rough estimate for the quantum decay time of a N\'eel-like state in a special ribbon geometry, namely
\begin{equation}
\tau_{\rm qd} \sim (L/5)^3\hbar/k_{\rm B}K \approx (L/5)^3 {\rm ps}.
\end{equation}

{\it Quantum-classical crossover.} Up to now, we have considered isolated ribbons. In order to determine, whether EM in an actual experiment is a quantum- or a classical phenomenon in which the spin polarization is zero or finite, respectively, we need to consider the environment of the ribbon, including  the measurement process. The simplest possible way of doing this is to collapse all the complicated system-environment interactions into one single environment timescale $\tau_{\rm env}$ on which quantum coherence within the ribbon is destroyed by the environment. One may also understand $\tau_{\rm env}$ as the typical time between successive measurements of the spin state of the ribbon by the environment. Such a measurement will be local and will certainly tend to destroy the subtle entanglement of the true ground state $|\Psi_0\rangle$, thereby preparing the ribbon in a classical non-entangled state, say $|\Psi_{\rm N}\rangle$. We have argued that $|\Psi_{\rm N}\rangle$ will decay on a timescale $\tau_{\rm qd}$ towards the quantum ground state. However, if the time $\tau_{\rm env}$ between two measurements is much shorter than $\tau_{\rm qd}$, the ribbon is prepared into {\it the same} classical state $|\Psi_{\rm N}\rangle$ over and over again and thus cannot decay. This phenomenon is known as the quantum Zeno effect \cite{misra_quantum_zeno_1977}. An ordinary bulk antiferromagnet behaves classically because $\tau_{\rm env}\ll \tau_{\rm qd}$. But in graphene ribbons $\tau_{\rm qd}$ can be tuned via the ribbon dimensions over a wide range, from below a ps up to $\micro$s and higher. The environment time $\tau_{\rm env}$ is expected in this range as well. Thus, graphene-based nanostructures are perfect candidates for the study of the delicate crossover between classical and quantum behavior.

{\it Conclusion.} Our study of edge magnetism clarifies the role of quantum fluctuations, which are usually neglected in mean-field approaches. They destroy the long-range spin correlations by forming {\it rung singlets}, an effect which can be seen clearly only in realistically large systems with $\gtrsim 10^4$ carbon atoms and with methods beyond mean-field. Furthermore, we have contrasted the classical- and quantum nature of edge magnetism, for which the timescale of decoherence by environment interactions is important. If decoherence is faster than the quantum dynamics, the system behaves classically and the notion of an edge spin polarization makes sense. In the opposite case the system is not a classical antiferromagnet but a subtle non-locally entangled spin-singlet. 
We have argued that the geometry determines the 
position on this quantum-classical crossover, for the study of
which graphene is thus a perfect playground. The feasibility of,
e.g., scanning tunneling spectroscopy to probe these different regimes
should be carefully considered.  
Furthermore, it will be important to explore in future research the consequences
of the strong (quantum) spin fluctuations for the usability of edge magnetism in spintronics applications.

\acknowledgements

We have profited from discussions with A. Harjun, C.~Honerkamp, C.~Koop, T. C. Lang, R.~Mazzarello, M.~Morgenstern, B. Trauzettel, and O. Yazyev. Financial support by the DFG under Grant WE 3649/2-1 is gratefully acknowledged, as well as the allocation of CPU time within JARA-HPC and from JSC J\"ulich.

\bibliography{refs}

\end{document}